\documentstyle[12pt]{article}
\catcode`\@=11
\def\makepreprititle{\par
  \begingroup
  \def\thefootnote{\fnsymbol{footnote}}
  \def\
@makefnmark{\hbox
  to 0pt{$^{\@thefnmark}$\hss}}
  \if@twocolumn
  \twocolumn[\@makepreprititle]
  \else \newpage
  \global\@topnum\z@
  \@makepreprititle \fi\thispagestyle{empty}\@thanks
  \endgroup
  \setcounter{footnote}{0}
  \let\makepreprititle\relax
  \let\@makepreprititle\relax
  \gdef\@thanks{}\gdef\@author{}\gdef\@title{}
  \gdef\@preprintnumber{}\gdef\@preprintdate{}\gdef\subtitle{}
  \let\thanks\relax}
\def\preprintnumber#1{\gdef\@preprintnumber{#1}}
\def\preprintdate#1{\gdef\@preprintdate{#1}}
\def\subtitle#1{\gdef\@subtitle{#1}}
\def\@makepreprititle{\newpage
{\def\baselinestretch{1}
  \begin{flushright} \@preprintnumber \par
  \@preprintdate \end{flushright} } \par
  \begin{center}
\vskip 1.5em
  {\LARGE \@title \par} \vskip 2.5em
  {\large \lineskip .5em
  \begin{tabular}[t]{c}\@author
  \end{tabular}\par}
  \vskip 1em {\large \@date} \end{center}
  \par
  \vfil}
\date{\sl Department of Physics, Tohoku University\\Sendai, 980 Japan}
\preprintdate{~}
\preprintnumber{~}
\subtitle{}
\def\abstract{\if@twocolumn
\section*{Abstract}
\else \normalsize
\begin{center}
{\bf Abstract\vspace{-.5em}\vspace{0pt}}
\end{center}
\quotation
\addtocounter{page}{-1}
\fi}
\def\endabstract{\if@twocolumn\else\endquotation\fi}
\catcode`\@=12
\preprintdate{July 1994}
\preprintnumber{TU-462\\DPSU-9403}
\title{Stability of Gauge Hierarchy \\ in Einstein Supergravity}
\author{Izumi Joichi,$^1$ Yoshiharu Kawamura$^2$ \\
and\\
 Masahiro Yamaguchi$^1$}
\date{\sl $^1$ Department of Physics, Tohoku University\\
Sendai, 980-77 Japan\\
$^2$ Department of Physics, Shinshu University\\
Matsumoto, 390 Japan}

\begin{document}
\makepreprititle
\begin{abstract}
In supersymmetric grand unified theories, the gauge hierarchy achieved
by a fine-tuning in the superpotential can be violated in the presence
of soft breaking terms.  We examine this problem in Einstein
supergravity with hidden-sector supersymmetry breaking.  We show that
the hierarchy is preserved if couplings of the hidden sector to
visible-sector fields in the superpotential satisfy a certain
requirement.
\end{abstract}

\newpage
\newcommand{\Kahler}{K\"ahler}
\newcommand{\la}{\langle}
\newcommand{\ra}{\rangle}

While supersymmetry (SUSY) is motivated as a solution of the gauge
hierarchy problem\cite{hierarchy,naturalness}, incorporation of soft
SUSY-breaking terms into the Lagrangian may spoil its stability.  A
well-known example is a model with a light singlet which has a
renormalizable coupling to the Higgs doublets in the superpotential
\cite{singlet1,singlet2}.  Then radiative corrections will generally
induce a large tadpole of the singlet shifting its vacuum expectation
value, which gives rise to  huge Higgs masses through the
renormalizable coupling.

 It is, however, less recognized that the gauge hierarchy can be
violated even without such a light singlet.  This will occur when the
smallness of the Higgs mass is not protected by any other symmetry but
the unbroken SUSY.  Once the SUSY breaking is turned on, a
(SUSY-breaking) Higgs mass would no longer remain at the weak scale.
This is especially problematic in a wide class of grand unified
theories (GUTs) where the supersymmetric mass of the Higgs doublets is
adjusted to be small by a fine-tuning of the parameters in the
superpotential \cite{SUSY-GUT}.  The fine-tuning of the SUSY invariant
part does not always achieve a small SUSY-breaking mass
simultaneously.  Indeed one can easily find an example of soft terms in
which the fine-tuning does not hold in the soft SUSY-breaking sector.
To keep the gauge hierarchy, the SUSY breaking terms should be
severely restricted.  In Ref.~\cite{KMY2}, it was shown that the large
mass vanishes if one adopts a certain ansatz for the soft
SUSY-breaking terms.  Furthermore it was pointed out that this ansatz
is stable against radiative corrections.  Indeed it can be derived from
the supergravity Lagrangian, if the so-called hidden sector completely
decouples from the visible sector in the superpotential.  However, the
most general form of the soft SUSY-breaking terms in which the
hierarchy is preserved was not clarified.

The purpose of this paper is to reconsider the stability of the gauge
hierarchy of the SUSY-GUTs in the framework of Einstein supergravity
\cite{SUGRA1,SUGRA2}.  Special attention will be paid to the effects
of the heavy fields to the masses of the light sector.  Since the
origin of the soft terms will be in the supergravity, our approach
 is natural and indeed will turn out to be transparent.  We
assume that the SUSY is spontaneously broken in the hidden sector
\cite{hidden}, though our argument may not depend on the specific SUSY
breaking mechanism.  We directly deal with the scalar potential of the
supergravity theory itself, without taking the flat limit first.  We
will show that the stability of the gauge hierarchy requires a
constraint on the couplings of the hidden sector to the visible
sector.  The ansatz of Ref.~\cite{KMY2}\ trivially satisfies it.
Relation of our argument to the entropy crisis problem
\cite{entropy-original}\ is also discussed briefly.

We begin by reviewing the scalar potential in the Einstein
supergravity.  It is specified by two functions,  the total K\"ahler
potential $G(\phi, \bar \phi)$ and the gauge kinetic function
$f_{\alpha \beta}(\phi)$ with $\alpha$, $\beta$  being indices of the adjoint
representation of the gauge group.   The former is a sum of
the K\"ahler potential $K$ and (the logarithm of) the superpotential
$W$
\begin{equation}
   G(\phi, \bar \phi)=K(\phi, \bar \phi) +M^{2}\ln |W (\phi) /M^{3}|^2,
\label{total-Kahler}
\end{equation}
where $M=m_{Pl}/\sqrt{8\pi}$ with $m_{Pl}$ being the Planck mass, and is
referred
to as the gravitational scale.
We have denoted scalar fields in the chiral multiplets by $\phi^{\kappa}$
and their complex conjugate by $\bar \phi^{\bar \kappa}$.  The
scalar potential is given by
\begin{equation}
   V= M^{2}e^{G/M^{2}} \hat{V}
     +\frac{1}{2} (Re f^{-1})_{\alpha \beta} \hat D^{\alpha}
\hat D^{\beta},
\label{scalar-potential}
\end{equation}
where
\begin{eqnarray}
  \hat{V} &=& G_\kappa (K^{-1})^{\kappa \bar \lambda} G_{\bar \lambda}-3M^{2},
\\
  \hat D^\alpha &=&  G_\kappa ( T^\alpha \phi)^\kappa
            = (\bar \phi T^\alpha)^{\bar \kappa} G_{\bar \kappa}.
\label{hatD}
\end{eqnarray}
Here $G_{\kappa}=\partial G/\partial \phi^\kappa$, $G_{\bar
\kappa}=\partial G/\partial \bar \phi^{\bar \kappa}$ etc, and
$T^\alpha$ are gauge transformation generators. Also in the
above, $(Re f^{-1})_{\alpha \beta}$ and  $(K^{-1})^{\kappa \bar \lambda}$
are the inverse matrices of $Re f_{\alpha \beta}$ and  $K_{\kappa \bar
\lambda}$ respectively, and summation over $\alpha$,... and  $\kappa$,... is
understood.  The last equality in Eq.~(\ref{hatD}) comes from the
gauge invariance of the K\"ahler potential.

Let us next summarize our assumptions on supersymmetry breaking.  The
gravitino mass $m_{3/2}$ is given by
\begin{equation}
   m_{3/2}= \langle Me^{G/2M^{2}} \rangle,
\label{gravitino}
\end{equation}
where $\langle \cdots \rangle$ denotes the vacuum expectation value
(VEV).  As a phase convention, it is taken to be real.  We
identify the gravitino mass with the weak scale.
The $F$-auxiliary fields of the chiral multiplets are
\begin{equation}
   F^\kappa =Me^{G/2M^{2}} (K^{-1})^{\kappa \bar \lambda} G_{\bar \lambda}.
\label{F}
\end{equation}
We require their VEVs should  satisfy
\begin{equation}
   \langle F^\kappa \rangle \leq O(m_{3/2}M).
\label{VEV-F}
\end{equation}
As will be seen shortly, the VEVs of the $D$-auxiliary fields become
 very small $\langle \hat D^\alpha \rangle \leq O(m_{3/2}^2)$
despite a naive expectation of order $m_{3/2}M$.  It follows from
Eq.~(\ref{VEV-F}) that
\begin{equation}
   \langle G_\kappa \rangle, \ \langle G_{\bar \kappa} \rangle
   \leq O(M)
\label{Gkappa}
\end{equation}
and
\begin{equation}
  \langle \hat V \rangle \leq O(M^{2}).
\label{Vhat}
\end{equation}
Note that    we allow the non-zero
vacuum energy $\langle V \rangle$ of order $m_{3/2}^2 M^2$ at this
level, which could be canceled by quantum corrections.
We also assume that derivatives of the K\"ahler potential with respect
to $\phi$ and $\bar \phi$ are at most of order unity (in the
units where $M$ is taken to be unity), namely
\begin{equation}
  \langle K_{\kappa_1 \cdots \bar{\lambda}_1 \cdots} \rangle \leq O(1).
\end{equation}
This will be justified if there is no strongly interacting sector.

The fields in the theory at hand are assumed to be classified into two
sectors. One is the heavy sector which has GUT scale mass, the other
is the light sector whose mass is typically in the weak scale.  Fields
in the (minimal) supersymmetric standard model belong to the light
sector.\footnote{We assume that the supersymmetric mass for the Higgs
doublets is of the weak scale at the vacuum after the
SUSY breaking is incorporated, which may be invalid in the presence of
a light singlet in the visible sector.  We disregard the case that a
light singlet scalar  exists in the visible sector.}
 In the following, we assume, for simplicity, that the mass scale of
the heavy sector is  identified with the gravitational scale $M$.
Following Ref.~\cite{HLW}, we further classify the heavy
sector into two, heavy complex and heavy real.  A heavy complex field
has a large mass from the superpotential. The pseudo scalar
counterpart of the latter is  the Nambu-Goldstone boson
absorbed into  a gauge boson in the GUT
symmetry breaking.

In order to discuss the masses of the light scalar bosons, it is
necessary to see consequences of the stationary conditions $\langle
V_\mu
\rangle = \langle \partial V /\partial \phi^\mu \rangle =0$.   From
Eq.~(\ref{scalar-potential}),  we find
\begin{eqnarray}
  V_\mu & =& M^{2}(e^{G/M^{2}})_\mu \hat V +M^{2}e^{G/M^{2}} \hat V_\mu
   +\frac{1}{2} (Re f^{-1})_{\alpha \beta, \mu} \hat D^\alpha \hat
D^\beta
\nonumber \\
  & & +(Re f^{-1})_{\alpha \beta} \hat D^\alpha (\hat D^\beta)_\mu
\nonumber \\
  &= &  G_\mu e^{G/M^{2}} \hat V
    + M^{2}e^{G/M^{2}}\{ G_{\mu \kappa} (K^{-1})^{\kappa \bar \lambda}
                       G_{\bar \lambda}
   -G_\lambda (K^{-1})^{\lambda \bar \lambda}K_{\kappa \bar \lambda
\mu} (K^{-1})^{\kappa \bar \kappa} G_{\bar \kappa}
\nonumber \\ & &   +G_\mu \}
 -\frac{1}{2} Re f_{\alpha \beta, \mu}D^\alpha D^\beta
       + D^\alpha (\bar \phi T^\alpha )^{\bar \kappa}
        K_{\bar \kappa  \mu},
\label{Vmu}
\end{eqnarray}
where
\begin{equation}
   D^\alpha =(Re f^{-1})_{\alpha \beta}\hat  D^\beta
\end{equation}
is the $D$-auxiliary field of the vector multiplet.

Let us now multiply $(T ^\alpha \phi)^\mu$ to the above, or project on
a heavy-real direction.  Using the identities derived from the gauge
invariance of the total K\"ahler potential
\begin{eqnarray}
   & & G_{\kappa \mu}(T^\alpha \phi)^\mu
   +G_\mu (T^\alpha )_\kappa ^\mu
   -K_{\kappa \bar \mu}(\bar \phi T^\alpha)^{\bar \mu}=0,
\\
   & & K_{\kappa \bar \lambda \mu} (T^\alpha \phi)^\mu
       +K_{\bar \lambda \mu } (T^\alpha)^\mu_\kappa
       -[G_{\bar \mu} (\bar \phi T^\alpha )^{\bar \mu}
         ]_{\kappa \bar \lambda}=0,
\end{eqnarray}
we obtain
\begin{eqnarray}
  \lefteqn{  V_{\mu} (T^\alpha \phi)^\mu =
       M^2 e^{G/M^2}(2+\hat V/M^2) \hat D^\alpha
       -F^\kappa \bar F^{\bar \lambda}
     (G_{\bar \mu} (\bar \phi T^\alpha )^{\bar \mu})_{\kappa \bar \lambda} }
\nonumber \\
 & & -\frac{1}{2} Re f_{\beta \gamma, \mu} (T^\alpha \phi)^\mu
       D^\beta D^\gamma
      +(T^\alpha \phi)^\kappa K_{\kappa \bar \lambda}
       (\bar \phi T^\beta)^{\bar \lambda} D^\beta .
\end{eqnarray}
Taking its VEV, we find
\begin{eqnarray}
\lefteqn{  0=
       m_{3/2}^2 (2+\la \hat V \ra /M^2) \la \hat D^\alpha \ra
       -\la F^\kappa \ra \la \bar F^{\bar \lambda} \ra
        \la  (G_{\bar \mu}
          (\bar \phi T^\alpha )^{\bar \mu})_{\kappa \bar \lambda} \ra }
\nonumber \\
 & & -\frac{1}{2}\la Re f_{\beta \gamma , \mu} (T^\alpha \phi)^\mu \ra
       \la D^\beta \ra \la  D^\gamma \ra
      +\frac{1}{2} M_{V}^{2 \alpha \beta} \la  D^\beta \ra,
\label{stationary-B}
\end{eqnarray}
where $M_{V}^{2 \alpha \beta}=2 (T^\alpha \phi)^\kappa K_{\kappa \bar
\lambda} (\bar \phi T^\beta)^{\bar \lambda}$ is, up to the
normalization due to the gauge coupling constants, the mass matrix of
the gauge bosons.  Recalling that $M_{V}^{2 \alpha \beta}$ are assumed
to be $O(M^2)$ for broken generators of the GUT symmetry, we conclude
\begin{equation}
  \la D^\alpha \ra \leq O(m_{3/2}^2) ,
\label{VEV-D}
\end{equation}
as the first three terms of Eq.~(\ref{stationary-B}) are already of
order $m_{3/2}^2 M^2$ or less.\footnote{For an unbroken generator, the
VEV of the $D$-term vanishes in the absence of the Fayet-Iliopoulos
$D$-term.} It is noteworthy that quite a similar equation
to (\ref{stationary-B}) is obtained for the case of a  non-linear
realization of the gauge symmetry.

We now return to $\la V_\mu \ra=0$ itself.  Taking the VEV of Eq.~(\ref{Vmu})
and using Eqs.~(\ref{F}), (\ref{Gkappa}), (\ref{Vhat}) and
(\ref{VEV-D}), we find
\begin{equation}
 \la M e^{G/2M^{2}} G_{\kappa \mu} \ra \la F^\kappa \ra \leq O(m_{3/2}^2 M)
\end{equation}
for any index $\mu$. Since $\la M e^{G/2M^{2}} G_{\kappa \mu} \ra=M_{\kappa
\mu}+O(m_{3/2})$ with $M_{\kappa \mu}$ being the supersymmetric mass
coming from the superpotential,  the above reads
\begin{equation}
M_{\kappa \mu} \la F^\kappa \ra \leq O(m_{3/2}^2 M).
\label{MkappamuF}
\end{equation}
Since we assume that $M_{\kappa \mu}$ is $O(M)$ for heavy complex fields,
we conclude
\begin{equation}
\la F^\kappa \ra \leq O(m_{3/2}^2)
\end{equation}
for them.  This will play an important role in the subsequent
arguments.
On the other hand,  Eq.~(\ref{MkappamuF}) implies that a hidden sector
field which has a large $F$-term of order $m_{3/2}M$ should be light
with the weak-scale mass.   Therefore, in our convention, the
hidden-sector field is contained in the light sector.

Note that, when the mass of the heavy complex field is below the
gravitational scale, a careful analysis tells us that
$\la F^\kappa \ra \leq O(m_{3/2}^2 \la \phi^\kappa \ra /
M_\kappa)$, where $M_\kappa$ is the mass of $\phi^\kappa$ from the
superpotential.
 Thus as far as  $\la \phi ^\kappa \ra \sim M_\kappa $,  the VEV of
its $F$-term is always small $\sim m_{3/2}^2$.

Now we would like to integrate out the heavy sector to obtain the
low-energy effective Lagrangian of the light scalar bosons.  The
procedure we should take consists of the three  parts: (1) We
calculate the VEVs of the derivatives of the potential so that we
write the potential as $V=\frac{1}{2} \la V_{k l} \ra \phi^k \phi^l
+\cdots$.  (2) When there exists mass mixing between the heavy and
light sectors, we should diagonalize them to correctly identify the
light and heavy fields. (3) Then we integrate out the heavy sector.
Practically we solve the heavy fields in terms of the light fields and
substitute their solutions to the potential.  Then we obtain the scalar
potential of the light fields only.

For our purpose to investigate the stability of the hierarchy, it is
sufficient to study the mass terms of the light fields.  When there is no
light-heavy mass mixing, the mass squared of the light scalar fields
are simply given by the VEVs of the second derivatives of the
potential.  From Eq.~(\ref{scalar-potential}), it follows that
\begin{eqnarray}
 \lefteqn{V_{\mu \nu}=
            \frac{\partial ^2 V}{\partial \phi^\mu \partial \phi^\nu}=}
\nonumber \\
  & & M^{2}(e^{G/M^{2}})_{\mu \nu} \hat V
     +M^{2}(e^{G/M^{2}})_\mu \hat V_\nu
     +M^{2}(e^{G/M^{2}})_\nu \hat V_\mu
     +M^{2}e^{G/M^{2}} \hat V_{\mu \nu}
\nonumber \\
   & &  +\frac{1}{2} (Re f^{-1})_{\alpha \beta, \mu \nu}
           \hat D^\alpha \hat D^\beta
        +(Re f^{-1})_{\alpha \beta, \mu }
           \hat D^\alpha (\hat D^\beta)_\nu
       +(Re f^{-1})_{\alpha \beta, \nu }
           \hat D^\alpha (\hat D^\beta)_\mu
\nonumber \\
     & &     +(Re f^{-1})_{\alpha \beta }
           \hat D^\alpha (\hat D^\beta)_{\mu \nu}
       +(Re f^{-1})_{\alpha \beta}
           (\hat D^\alpha)_\mu (\hat D^\beta)_\nu.
\end{eqnarray}
 From Eqs.~(\ref{Vhat}), (\ref{Gkappa}) and (\ref{VEV-D}), we get
\begin{eqnarray}
\la V_{\mu \nu} \ra &=&\la M^{2}(e^{G/M^{2}})_{\mu \nu} \hat V\ra
        +\la M^{2}e^{G/M^{2}} \hat V _{\mu \nu} \ra
	+\la (Re f^{-1})_{\alpha \beta} \ra
	      \la (\hat D^\alpha)_\mu \ra  \la (\hat D^\beta)_\nu \ra
\nonumber \\
     & &  +O(m_{3/2}^2),
\label{Vmunu}
\end{eqnarray}
where we have used $\langle \hat V_{\mu} \rangle \leq O(1)$.
Similarly the chirality-conserving mass terms are found to be
\begin{equation}
     \la V_{\mu \bar \nu} \ra
    = \la M^{2}e^{G/M^{2}} \hat V_{\mu \bar \nu} \ra
     +\la (Re f^{-1})_{\alpha \beta} \ra \la (\hat D^\alpha )_\mu \ra
                                  \la (\hat D^\beta )_{\bar \nu} \ra
     +O(m_{3/2}^2).
\label{Vmubarnu}
\end{equation}

Let us first evaluate Eq.~(\ref{Vmunu}). In the minimal supersymmetric
standard model, this contains the mixing mass term for the two Higgs
doublets. An inspection shows
\begin{eqnarray}
  \hat V_{\mu \nu}&=& G_{\kappa \mu \nu }
                      (K^{-1})^{\kappa \bar \lambda}
                      G_{ \bar \lambda}
               +2G_{\mu \nu}
\nonumber \\
          & & -G_{\mu \kappa}(K^{-1})^{\kappa \bar \lambda}
            K_{\bar \lambda \rho \nu}(K^{-1})^{\rho \bar \sigma}
            G_{\bar \sigma}
           -G_{\nu \kappa}(K^{-1})^{\kappa \bar \lambda}
            K_{\bar \lambda \rho \mu}(K^{-1})^{\rho \bar \sigma}
            G_{\bar \sigma}
\nonumber \\
        & & +G_{\kappa} (K^{-1})^{\kappa \bar \lambda} _{, \mu \nu}
          G_{\bar \lambda}
         -G_{\kappa} (K^{-1})^{\kappa \bar \lambda} K_{\bar \lambda
\mu \nu}.
\end{eqnarray}
Thus Eq.~(\ref{Vmunu}) reads
\begin{eqnarray}
\la V_{\mu \nu} \ra &=& \la Me^{G/2M^{2}}(G_{\mu \nu \kappa}
  	-G_{\mu \lambda} (K^{-1})^{\lambda \bar \rho}
         K_{\bar \rho \kappa \nu}
        -G_{\nu \lambda} (K^{-1})^{\lambda \bar  \rho}
         K_{\bar \rho \kappa \mu} ) \ra \la F^\kappa \ra
\nonumber \\
  & +& (2+\la \hat V \ra /M^{2})m_{3/2} \la Me^{G/2M^{2}} G_{\mu \nu} \ra
   +\la (Re f^{-1})_{\alpha \beta} \ra
    \la (\hat D^\alpha)_\mu \ra  \la (\hat D^\beta )_\nu \ra
\nonumber \\
& +& O(m_{3/2}^2).
\label{VEV-Vmunu}
\end{eqnarray}
When both $\phi^\mu$ and $\phi^\nu$ belong to the light sector,
$\la G_{\mu \kappa} \ra$, $\la G_{\nu \kappa } \ra  \leq O(1)$
for any  $\phi^\kappa$ and $\la (\hat D^\alpha)_\mu \ra $, $ \la (\hat
D^\beta )_\nu \ra \leq O(m_{3/2})$ by the very definition of the light
fields.  Thus we
find
\begin{equation}
\la V_{\mu \nu} \ra =\la Me^{G/2M^{2}} G_{\mu \nu \kappa} \ra
                    \la F^\kappa \ra +O(m_{3/2}^2).
\end{equation}

For $\la V_{\mu \bar \nu} \ra$, we  only give a
result
\begin{eqnarray}
\la V_{\mu \bar \nu} \ra &=& \la Me^{G/2M^{2}} G_{\mu \kappa} \ra
                           \la (K^{-1})^{\kappa \bar \lambda} \ra
                          \la Me^{G/2M^{2}} G_{\bar \lambda \bar \nu}
\ra
\nonumber \\
      & &- \la Me^{G/2M^{2}} G_{\mu \kappa} \ra \la (K^{-1})^{\kappa \bar
\lambda} K_{\bar \lambda \bar \nu \rho} \ra \la F^\rho \ra
\nonumber \\
  & & - \la Me^{G/2M^{2}} G_{\bar \nu \bar \kappa} \ra \la (K^{-1})^{\bar
\kappa
\lambda} K_{ \lambda  \mu \bar \rho} \ra \la \bar F^{\bar \rho} \ra
\nonumber \\
  & &      +\la (Re f^{-1})_{\alpha \beta} \ra
                  \la (\hat D^\alpha )_\mu \ra
                  \la (\hat D^\beta )_{\bar \nu} \ra
\nonumber \\
& & +O(m_{3/2}^2).
\label{VEV-Vmubarnu}
\end{eqnarray}
We can see this is always of order $m_{3/2}^2$ for light fields
$\phi^\mu $ and $\phi^{\bar \nu}$.

To summarize, we find the condition that the gauge hierarchy survives
 after integrating out the heavy sector to be
\begin{equation}
    \la Me^{G/2M^{2}}G_{\mu \nu \kappa} \ra \la F^\kappa \ra \leq
O(m_{3/2}^2).
\label{GmunukappaF}
\end{equation}
Note that, if there are light-heavy mixing mass terms, we have to
diagonalize the mass matrix to identify the heavy and light fields
correctly. From Eqs.~(\ref{VEV-Vmunu}) and (\ref{VEV-Vmubarnu}), we find
that these mixing mass terms are at most $O(m_{3/2}M)$, therefore
their contributions to the mass terms of the light sector are of
order $m_{3/2}^2$ or less.  Hence Eq.~(\ref{GmunukappaF}) applies also to
this case.

We are now ready to discuss the contribution of the heavy sector to
the above condition (\ref{GmunukappaF}).  In many of the SUSY-GUT
models, there exists a coupling of order unity $\la
Me^{G/2M^{2}}G_{\mu \nu \kappa} \ra$ for a heavy complex field
$\phi^\kappa$.  For example, in the minimal $SU(5)$, the light Higgs
multiplets couple to the $SU(5)$ adjoint Higgs, which would embarrass
the hierarchy in the context of global SUSY\cite{KMY2}.  However, in
the supergravity, we showed that $\la F^\kappa \ra$ for the heavy
complex field is always small $\sim m_{3/2}^2$ due to the stationary
condition Eq.~(\ref{MkappamuF}), and hence this coupling does not
upset the requisite gauge hierarchy. On the other hand, heavy real
fields can have large $\la F^\kappa \ra$ of $O(m_{3/2}M)$. 	But in
this case, the gauge invariance of $G$ gives
\begin{equation}
    G_{\kappa \mu \nu}(T^\alpha \phi)^\kappa
   +G_{\kappa \mu} (T^\alpha )^\kappa _\nu
   +G_{\kappa \nu} (T^\alpha )^\kappa _\mu
   -K_{\bar \kappa \mu \nu}(\bar \phi T^\alpha)^{\bar \kappa} =0,
\end{equation}
which implies
\begin{equation}
    \la G_{\kappa \mu \nu}(T^\alpha \phi)^\kappa \ra \leq O(1),
\end{equation}
for $\mu, \nu$ light. Therefore the coupling
$\la Me^{G/2M^{2}}G_{\kappa \mu \nu} \ra$ is suppressed to $\leq O(m_{3/2}/M)$
and again it does not spoil the condition (\ref{GmunukappaF}).

We thus conclude  that the heavy sector does not give any large SUSY
breaking mass terms which destabilize the weak scale.  In the absence
of a light singlet in the visible sector,  the requirement
(\ref{GmunukappaF}), therefore, gives a constraint on the  couplings of the
hidden sector to the visible sector.

To illustrate this point,  we now consider a toy
model with the superpotential
\begin{equation}
   W= h(z) +(M_H (z)-\lambda (z) \Sigma) H \bar H + \cdots,
\end{equation}
where $H$ and  $\bar H$ are light Higgs multiplets, $\Sigma$ is  a heavy field
which has
a large VEV $\la \Sigma \ra =O(M)$, and $z$ is a hidden-sector scalar
field responsible for the SUSY breaking.   The mass term $M_H(z)$ and
the Yukawa coupling $\lambda (z)$ may depend on the hidden field $z$.
Suppose that the supersymmetric mass of $H$, $\bar H$ are at the weak
scale by a fine-tuning
\begin{equation}
   \la M_H (z)-\lambda (z) \Sigma \ra =O(m_{3/2}).
\end{equation}
Now the effective Yukawa  coupling of $H \bar H z$ is given by
\begin{equation}
  \la M e^{G/2M^2} G_{H \bar H z} \ra
  =\la e^{K/2M^2}(M_H^\prime (z) -\lambda^\prime (z) \Sigma) \ra +O(m_{3/2}/M),
\label{Yukawa}
\end{equation}
where the prime means the derivative with respect to $z$.
Eq.~(\ref{GmunukappaF}) requires $\la M_H^\prime (z) -\lambda^\prime (z)
\Sigma \ra$ is of order $m_{3/2}/M$ or less, since $\la F^z \ra$ is of
order $m_{3/2}M$.  However if the $z$ dependence of $M_H(z)$ is
different from that of $\lambda (z)$, then Eq.~(\ref{Yukawa}) will
generally become of order unity.   A form of the superpotential which
satisfies the above requirement is
\begin{equation}
    W=h(z) +g(z)f(H, \bar H, \Sigma).
\end{equation}
then $M_H(z)-\lambda(z) \Sigma$ is factorized to $g(z)  (M_H-\lambda
\Sigma)$. Here $M_H$ and $\lambda$ are constants, which are chosen so
that $\la M_H-\lambda \Sigma \ra =O(m_{3/2})$.
Then
\begin{equation}
   \la  M_H^\prime (z)-\lambda^\prime (z) \Sigma \ra
  = \la g^\prime (z) \ra \la M_H-\lambda \Sigma  \ra
  =O(m_{3/2}/M),
\end{equation}
which meets our requirement.   Note that when the hidden sector
completely decouples from
 the visible sector,  $g(z)$ is a constant
and the above equation is trivially satisfied.

Finally we would like to make a comment on Eq.~(\ref{MkappamuF}).  As
we mentioned above,  this equation implies that the ``Polonyi field''
with $\la F \ra =O(m_{3/2}M)$ should have a weak-scale mass.
This fact is related to the notorious entropy-crisis
problem \cite{entropy-original}, a drawback of the
usual hidden-sector SUSY-breaking scenario that the coherent
oscillation of the light Polonyi field  and its late decay will cause
huge entropy production.    Note that in deriving
Eq.~(\ref{MkappamuF}),  we did not use the  vanishing cosmological
constant condition and thus the lightness of the Polonyi field is not
a consequence of the zero cosmological constant,  but a generic
property of the $F$-term supersymmetry breaking.   See
Refs.~\cite{entropy-new,MoroiYanagida} for recent discussions on this
issue.

In this paper, we have discussed the stability of the gauge hierarchy
using the potential of the Einstein supergravity.  Since this
potential can be regarded as the tree-level potential, one may ask how
loop effects will change our argument.  To study radiative corrections of the
Einstein supergravity is a challenging issue and was partially
discussed, for example, in Ref. \cite{Gaillard}, which showed that the
most  of the logarithmic divergent part to the scalar potential
is  absorbed by the redefinition of the
\Kahler\ potential.  The effects of the renormalization of the \Kahler\
potential are already incorporated in our argument, since we do not
assume a specific form of it.  On the other hand, the whole effective
potential of the theory will be beyond the Einstein supergravity.  We
speculate here that a similar argument will apply to this case, and
the VEV of the $F$-term for a heavy complex will  remain small.  We
hope it will be discussed elsewhere \cite{JY}.

\section*{Acknowledgment}
We would like to thank H.~Murayama for stimulating discussions and
collaboration at an early stage of this work.  We are also grateful to
K.~Inoue and T.~Yanagida for useful discussions.


\end{document}